\documentclass[a4paper,pre,showpacs,aps,amsmath,amssymb,twocolumn]{revtex4}
\usepackage{graphicx}
\usepackage{graphics}
\usepackage{amsmath}
\usepackage{dcolumn}
\usepackage{amssymb}
\usepackage{bm}

\begin{document}
\title{Reverse Rotations in the Circularly-Driven Motion of a Rigid Body}
\author{Fernando Parisio}
\affiliation{Departamento de F\'{\i}sica, Universidade Federal de Pernambuco, 50670-901, Recife, PE, Brazil and 
Instituto de F\'{\i}sica, Universidade Federal de Alagoas, 57072-970, Macei\'o, AL, Brazil}

\begin{abstract}
We study the dynamical response of a circularly-driven rigid body, focusing on the
description of intrinsic rotational behavior (reverse rotations). The model system we
address is integrable but nontrivial, allowing for qualitative and quantitative analysis.
A scale free expression defining the separation between possible spinning regimes is obtained.
\end{abstract}
\maketitle

Nontrivial effects may arise in the realm of integrable classical
mechanics. The most recent manifestation of this fact is the
refreshed interest in the dynamics of sliding bodies
\cite{nature}. In particular, the work by Farkas {\it et al}
\cite{zeno}, an investigation on the frictional
coupling between translational and rotational motions, has
deserved attention, being followed by a dozen of papers on related
topics \cite{sliding}. Most of these previous work address the
influence of friction in the free (not driven) dynamics of disks.
In the present paper we deal, in a sense, with a complementary problem, namely,
that of frictionless dynamics of a forced rigid body (RB). More specifically,
we study the dynamical response of RB's submitted to rotational
driving forces (in a way that will become clear below), focusing
on the appearance of {\it reverse rotations}. Generally speaking,
a reverse rotation occurs when a system, or part of it,
is forced to rotate counterclockwise and its intrinsic angular
degree of freedom develops a clockwise motion, or vice-versa. This
terminology has been recently used in the literature to designate
unexpected rotational behavior of a cylinder inside a rotating
drum filled with a viscous fluid \cite{reverse}. Related, though more
intricate phenomena have been reported in the chaotic response of
a parametrically excited pendulum \cite{pendulum}, and in tests of
printing machinery of journals \cite{tribology}. Therefore,
reverse rotations are quite a general behavior in diverse physical systems.
In what follows we propose a simple mechanical model to investigate this effect
in an analytical way.

\begin{figure}
\includegraphics[width=4.2cm,angle=0]{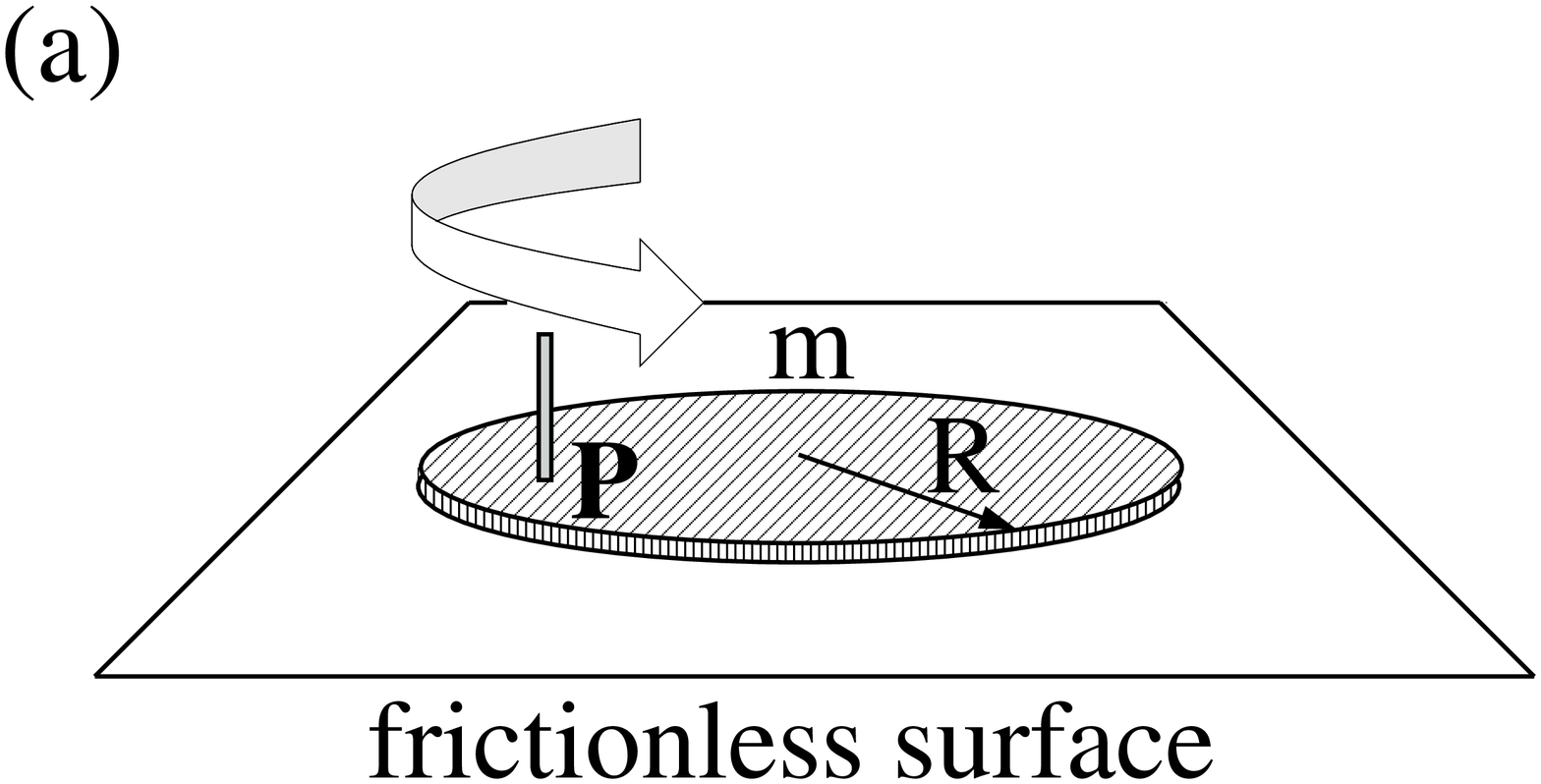}
\includegraphics[width=4.2cm,angle=0]{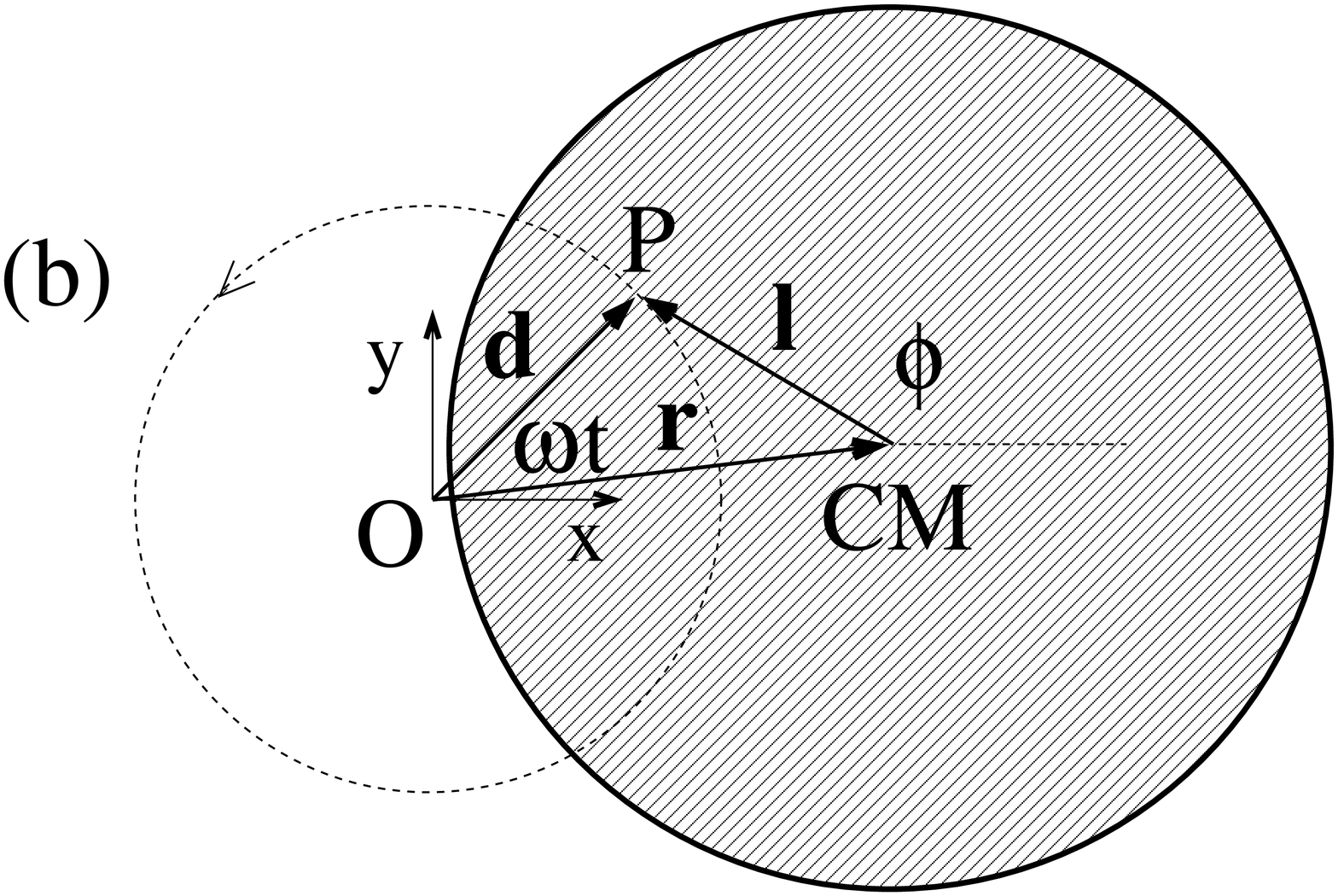}
\caption{(a) Circularly-driven motion of a disk on a frictionless horizontal surface.
(b) The point $P$, located by the vector ${\bf d}$, describes a circular path around
the origin $O$. The vector ${\bf r}$ gives the position of the center of mass ($CM$)
and ${\bf l}$ is the vector that
connects $CM$ to $P$.}
  \label{fig1}
\end{figure}

Our model system is schematically shown in Fig. \ref{fig1}(a). It
consists of a uniform disk of mass $m$ and radius $R$ resting on a
horizontal frictionless surface. We consider a disk only for convenience,
the following reasoning is valid for an arbitrary RB (this point shall
be discussed
at the end of the manuscript). The system is then submitted to
an external horizontal force ${\bf F}$, provided by a driving
apparatus, through a thin rod attached to a fixed point ($P$) on
the disk, around which the whole body can rotate freely. The
driving apparatus takes the disk from rest and makes the point $P$
follow a uniform circular trajectory of radius $d$ around a
fixed origin ($O$) with angular frequency $\omega$ (see Fig.
\ref{fig1}(b)). For definiteness we assume the rotation to be
counterclockwise, and without loss of generality we use a
coordinate system such that the point $P$ lies in the positive
$x$-axis at $t=0$. For later times we denote the position vector
of $P$ by ${\bf d}$ and the vector locating the center of mass
($CM$) by ${\bf r}$. Since the disk is assumed to be perfectly
rigid, $P$ is always a distance $l$ apart from $CM$. The relative
position of these two points is given by the vector ${\bf l}$, as
shown in Fig. \ref{fig1}(b). Finally, the angle between the $x$-axis
and the line connecting $CM$ and $P$ is denoted by $\phi$. The
variables ${\bf r}$ and $\phi$ completely specify the
position of the disk.

The main goal of this work is to answer the following question:
Which range of initial angles $\phi_0$ and parameters $m$,
$\omega$, $R$, $l$, and $d$, lead to a clockwise rotation of the
disk around $P$, if this is possible at all? The well-posedness
of this question shall become evident soon. Since the point $P$
itself is being forced in the counterclockwise direction, such a dynamical response
characterizes a reverse rotation (decreasing $\phi(t)$ on
average).

According to our definitions, we have
${\bf d}=d \cos(\omega t) \;{\bf \hat{ x}}+ d \sin(\omega t)\; {\bf \hat{ y}}$,
where $d$ can take any fixed value in the interval $(0,\, \infty)$. The vector
${\bf l}$ linking $CM$ to $P$ is given by
${\bf l}=l \cos \phi \;{\bf \hat{ x}}+ l \sin \phi\; {\bf \hat{ y}}$,
with $0<l\le R$. The position of $CM$ is denoted by ${\bf r}=x \;{\bf \hat{ x}}+ y\; {\bf \hat{ y}}$.
These three vectors must satisfy ${\bf r}+{\bf l}={\bf d}$,
which provides two holonomic constraints:
$x=d\cos(\omega t)-l\cos \phi$, and $y=d \sin(\omega t)-l\sin \phi$.

One initially cope with the problem without being concerned
with the question of how the disk came from rest to motion. This point
shall be addressed later. Suppose that ${\bf F}_c$, unknown {\it
a priori}, is the constraint force that keeps the circular
trajectory of $P$. This force is assumed to be provided by a
robust apparatus in the sense that the circular path is not
affected by the inertia of the disk. The equations of motion for
the $CM$ degrees of freedom read
$F_{c,x}=m\ddot{x}$,  $F_{c,y}=m\ddot{y}$.
The Newton's Second Law for the angular coordinate is given
by ${\bf \tau}_c= {\bf l} \times {\bf F}_c=I_{CM} \ddot{\phi}\, {\bf \hat{z}}$, i.e.,
\begin{equation}
ml(\cos \phi \;\ddot{y}-\sin \phi \;\ddot{x})=I_{CM} \ddot{\phi}\;,
\label{Eq.-phi}
\end{equation}
where $I_{CM}$ is the inertia moment of the disk with respect to the center of mass \cite{comment} .
By using the holonomic constraints we find $\ddot{x}=l \sin \phi \;\ddot{\phi}+l
\cos\phi\; \dot{\phi}^2- d \omega^2 \cos(\omega t)$ and $\ddot{y}=-l \cos \phi\;
\ddot{\phi}+l \sin \phi\; \dot{\phi}^2-d \omega^2 \sin(\omega t)$, one can, therefore,
decouple the angular equation of motion, which becomes:
$ml[ d\omega^2(\sin \phi \cos (\omega t)-\cos \phi \sin (\omega t))-l \ddot{\phi}]=
I_{CM} \ddot{\phi}$,
or
\begin{equation}
\ddot{\phi}-\frac{mld\omega^2}{I_P}\sin(\phi-wt)=0\;,
\label{Eq.-phi2}
\end{equation}
where $I_P=I_{CM}+ml^2$. As expected, the uniform plane rotation generates an effective
gravity.
We now proceed to the following change of variables:
$\theta=\phi-\omega t+\pi$,
implying $\dot{\theta}=\dot{\phi}-\omega$ and $\ddot{\theta}=\ddot{\phi}$.
This change transforms Eq. (\ref{Eq.-phi2})
into the simple pendulum equation $\ddot{\theta}+\frac{mld}{I_P}\omega^2\sin\theta=0$.
Thus, we see that the time dependence of $\phi$, the physically relevant variable, is given by a combination of pendular
and uniform motions (plus a constant additional factor),
\begin{equation}
\phi(t)=\theta_{pendulum}(t)+\omega t -\pi\;.
\label{sol}
\end{equation}
Note that solution (\ref{sol}) automatically yields $x(t)$, $y(t)$, and ${\bf F}_c(t)$.
It also implies that we have a hidden constant of the motion corresponding to
the mechanical energy of the auxiliary pendulum (we shall use this terminology
to refer to the pendular term in the solution (\ref{sol})) ${\cal E}=\frac{1}{2}I_P
\dot{\theta}^2+2mdl\omega^2\sin^2(\theta/2)$, where the potential energy is set
to zero in its lower position. In terms of initial conditions of the original variable we have
\begin{equation}
{\cal E}=\frac{1}{2}I_P (\dot{\phi}_0-\omega)^2+2mdl\omega^2\cos^2(\phi_0/2)\;.
\label{constant}
\end{equation}
By inspecting solution (\ref{sol}) we note that reverse rotations of the disk are possible
only when $\theta_{pendulum}$ describes a motion of rotation instead of libration, that is,
the mechanical energy of the auxiliary pendulum must satisfy ${\cal E}>2mdl\omega^2$ in
negative pendular cycles. Relation (\ref{constant}) together with the equality
${\cal E}=2mdl\omega^2$ define the curves that separate libration and rotation
of variable $\theta$ in the space of initial conditions $(\phi_0, \, \dot{\phi_0})$.
One finds that initial conditions in between the curves
\begin{equation}
\nonumber
\dot{\phi}_0=\omega \left[1 \pm 2\sqrt{\frac{mdl}{I_P}}\sin( \phi_0/2)  \right]\;
\label{libration}
\end{equation}
lead to libration of $\theta$, and thus, to normal rotation of the disk.
Of course, not all points that lie outside this region lead to reverse behavior. The pendular
rotation must be clockwise with an angular frequency larger than $\omega$, in order to make
$\phi(t)$ a decreasing function of $t$, on average. The condition for reverse rotation then
reads $T=\sqrt{\frac{8I_P}{\cal E}}K\left(\omega \sqrt{\frac{2mdl}{\cal E}}\right)<
\frac{2\pi}{\omega}$,
for clockwise pendular cycles of period $T$, where $K$ denotes the complete elliptic
function of first kind. In the regions where $T$ equals $2\pi/\omega$, the constant ${\cal E}$
satisfies the transcendental equation
\begin{equation}
K\left(\omega \sqrt{\frac{2mdl}{\cal E}}\right)=\frac{\pi}{\omega}\sqrt{\frac{\cal E}{2I_P}}\;.
\label{trans}
\end{equation}
Using this prescription and Eq. (\ref{constant}) we manage to select the initial conditions
$(\phi_0, \, \dot{\phi_0})$ that lead to reverse and normal rotations. These regimes are
separated by the curve
\begin{equation}
\dot{\phi}_0=\omega -\sqrt{\frac{2\tilde{\cal E}}{I_P}-\frac{4mdl\omega^2}{I_P}\cos^2( \phi_0/2)}\;,
\label{rr}
\end{equation}
where $\tilde{\cal E}$ stands for the solution of Eq.
(\ref{trans}). The root with a plus sign was discarded because it
is related to positive pendular rotations. We name the above
defined curve a synchronization line because initial conditions on
it develop neither normal nor reverse rotation, for the angular
frequency of $\theta$ coincides with $\omega$ and, therefore, the
time evolution of $\phi$ averages to a constant value.

At this point we address the problem with static initial conditions. This
is the most interesting situation, since it is expected that high
enough clockwise initial velocities $\dot{\phi}_0$ trivially
lead to reverse rotations. We note, however, that
$\dot{x}$, $\dot{y}$, and $\dot{\phi}$ never vanish
simultaneously, and, thus, the kinetic energy of the disk
%
${\cal K}
=m\dot{x}^2/2+m\dot{y}^2/2+I_{CM}\dot{\phi}^2/2
=md^2\omega^2/2+\left(I_P \omega/2+mdl \omega -{\cal E}/\omega \right)\dot{\phi}
-I_P\dot{\phi}^2/2+I_P\dot{\phi}^3/2\omega\;,$
%
is nonzero for all times. It is clear that the force ${\bf F}_c$
alone is not compatible with static initial conditions. In order
to encompass these conditions, we assume that an impulsive force
acts on the disk, taking it from rest to motion in a time scale
much shorter than any other in the problem, e. g. $2\pi/\omega$.
This is a realistic assumption when we have a table top engine
driving a light body. More explicitly, we suppose that the force
can be split into two parts
\begin{equation}
{\bf F}=\left\{
\begin{array}{c}
{\bf F}_0 \;\;  \mbox{for} \;\;t \in [0^-,0^+]\\
{\bf F}_c\;\; \mbox{for} \;\;t>0^+
\end{array}
\right.
\end{equation}
where ${\bf F}_0$ denotes the impulsive force that acts during an
arbitrarily small time interval centered at $t=0$. In the limiting
case we have a delta function, which we initially write in generic
form as ${\bf F}_0=\delta (t)(\alpha \,{\bf \hat{ x}}+ \beta \,
{\bf \hat{ y}})$, where $\alpha$ and $\beta$ have dimension of
momentum. These constants are to be determined by the motion we
{\it know} the force ${\bf F}_0$ causes to the disk. More
specifically, we know that, immediately after its application,
to fit the ${\bf F}_c$ prescription,
the point $P$ must acquire a velocity
\begin{equation}
{\bf v}_P(0^+)=\omega d \, {\bf \hat{ y}}\; .
\label{vp}
\end{equation}
The equations of motion in the infinitesimal interval $[0^-,0^+]$ are $m
\ddot{x}=\alpha \delta(t)$, $m \ddot{y}=\beta \delta(t)$, and $l \cos \phi
\beta \delta(t)-l \sin \phi \alpha \delta(t)=I_{CM}\ddot{\phi}$.
Integrating one obtains the velocities soon after the application of ${\bf F}_0$
\begin{equation}
\dot{x}(0^+)=\frac{\alpha}{m}\;,\;\; \dot{y}(0^+)=\frac{\beta}{m}\;,\;\;
\dot{\phi}(0^+)=\frac{l}{I_{CM}}(\beta \cos \phi_0-\alpha \sin \phi_0)\;,
\label{0+}
\end{equation}
where, here, the static initial conditions were employed:
$\dot{x}(0^-)=0$, $\dot{y}(0^-)=0$, and $\dot{\phi}(0^-)=0$. We
also used $\phi(0^-)=\phi(0^+)=\phi_0$ (since the impulsive force
causes no discontinuity in coordinates). The velocity of $P$ at any
time is given by ${\bf v}_P={\bf \dot{r}}+{\bf \dot{l}}=
(\dot{x}-l\dot{\phi}\sin \phi)\, {\bf \hat{
x}}+(\dot{y}+l\dot{\phi}\cos \phi) \, {\bf \hat{ y}}$. Therefore,
the initial velocity as a function of $\alpha$ and $\beta$ is
\begin{eqnarray}
\nonumber
{\bf v}_P(0^+)=\left[ \frac{\alpha}{m}-\frac{l^2}{I_{CM}}\sin
\phi_0(\beta \cos \phi_0 - \alpha \sin \phi_0) \right]\, {\bf \hat{x}}\\
\nonumber
+ \left[ \frac{\beta}{m}+\frac{l^2}{I_{CM}}\cos
\phi_0(\beta \cos \phi_0 - \alpha \sin \phi_0) \right]\, {\bf \hat{y}} \;.
\end{eqnarray}
By applying the consistency condition (\ref{vp}) we obtain a pair of equations,
involving $\alpha$ and $\beta$, whose solution is
\begin{equation}
\nonumber
\alpha = \frac{\omega d m^2 l^2}{I_P}\sin \phi_0 \cos \phi_0\;,\;\; \beta=
md \omega\left( 1-\frac{m l^2}{I_P}\cos^2 \phi_0\right)\;.
\end{equation}
We then found the impulsive force that is consistent with the
subsequent evolution of the system. Substituting the above
results in the last equation in (\ref{0+}) we obtain a quite
simple relation between the initial angle $\phi_0$ and the initial
velocity $\dot{\phi}_0$ (where we suppress the argument $0^+$).
Given the initial angle of the static disk, the angular velocity
it acquires immediately after the driving apparatus is turned on
is
\begin{equation}
\dot{\phi}_0=\frac{mdl\omega}{I_P}\cos \phi_0\;.
\label{cc}
\end{equation}
\begin{figure}
\includegraphics[width=4.5cm,angle=-90]{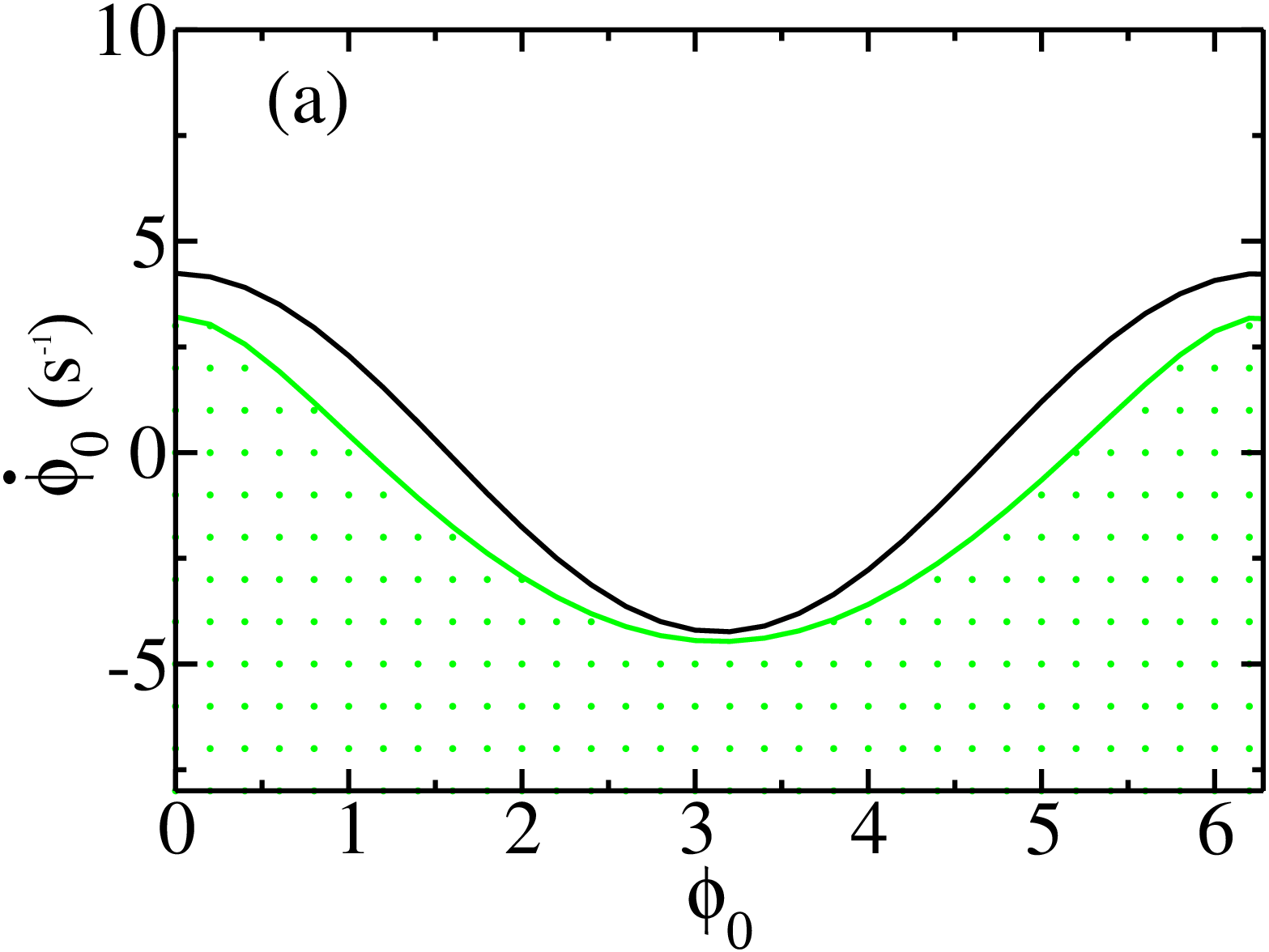}
\includegraphics[width=4.5cm,angle=-90]{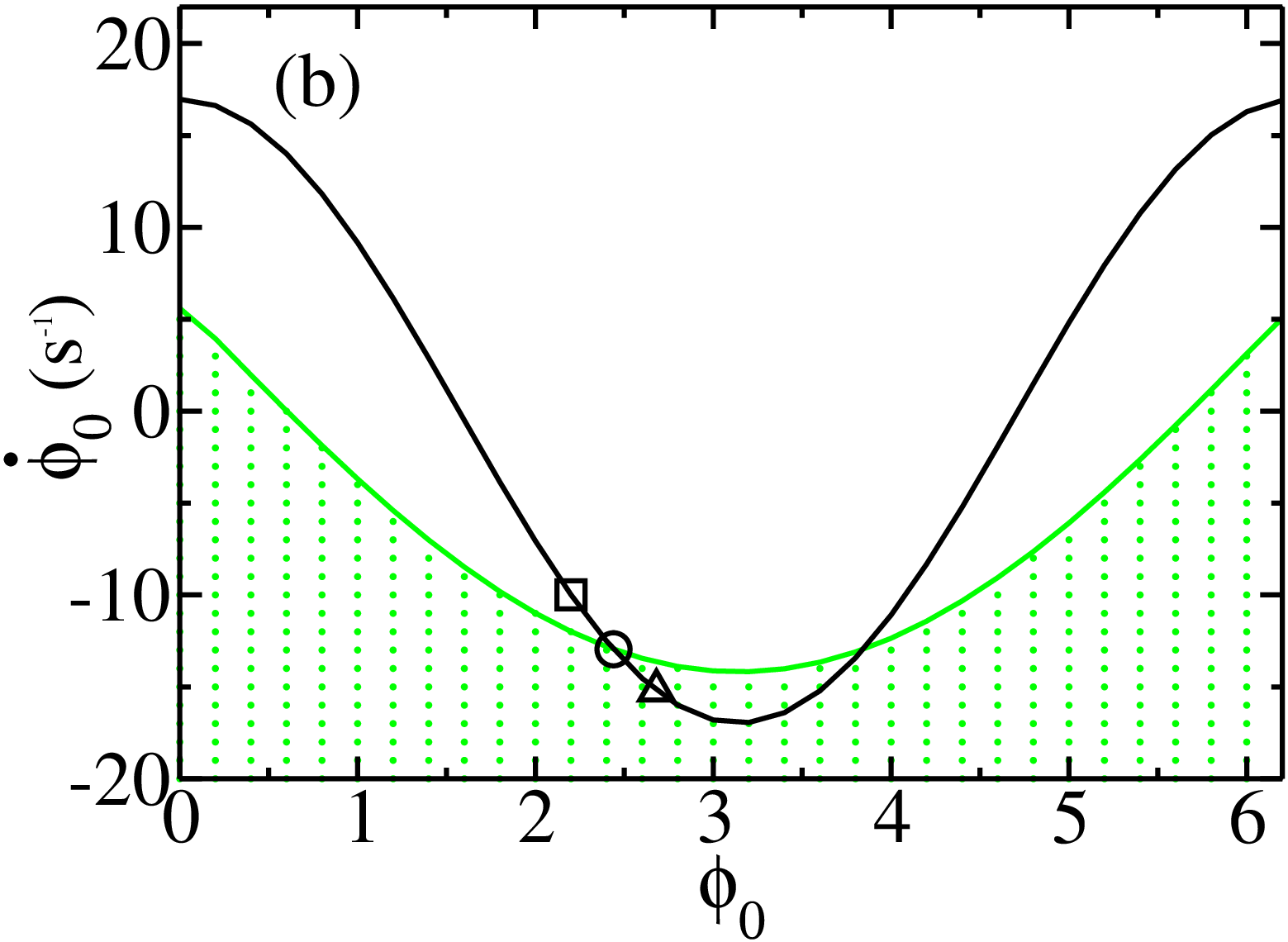}
\caption{The dotted regions represent initial conditions leading to
reverse rotations. We set $d=10$ cm in (a) and $d=40$ cm
in (b). The grey and black lines are the synchronization and initial
condition curves, respectively. In (a) no reverse rotations can
develop since the two lines do not cross. In (b) reverse behavior is
possible for an interval centered in $\phi_0=\pi$.}
  \label{fig2}
\end{figure}
\begin{figure}
\includegraphics[width=3.5cm,angle=-90]{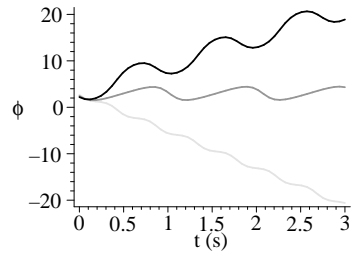}
\caption{Evolution of the initial conditions indicated in Fig. 2 (b):
square (black), circle (grey), and triangle (light grey).}
  \label{fig3}
\end{figure}
Only pairs $(\phi_0,\dot{\phi}_0)$ related through this expression
are valid initial conditions. We are now in position to show the
relevant regions and curves in the space of initial conditions. In
Fig. \ref{fig2}(a) we show the normal (blank) and reverse (dotted)
regions in the space $(\phi_0,\dot{\phi}_0)$, separated by the
synchronization line (grey curve). The black
line represents the possible initial conditions as given by
(\ref{cc}). We used the following parameters: $\omega=6$ rad/s
($\nu \sim 1$ Hz), $m=100$ g, $R=10$ cm, $l=7$ cm, and $d=10$ cm.
For these values, the constant of the motion that satisfies
Eq. (\ref{trans}) is $\tilde{\cal E} \approx 0.0543$ J. We see
that, for these parameters, no reverse rotations can occur (the
black curve do not reach the dotted region). In contrast, if we
set $d=40$ cm, keeping the other parameters (leading to $\tilde{\cal E}
\approx 0.2017$ J), we obtain the result displayed in Fig.
\ref{fig2}(b), where it is clear (the grey and black curves cross
each other) that reverse rotations take place for an interval of
initial angles centered in $\phi_0=\pi$. The three initial
conditions marked with a square, a circle, and a triangle
represent the possible regimes: normal rotation, synchronization,
and reverse rotation. The time evolution of $\phi$ for these three
conditions is shown in Fig. \ref{fig3}. The corresponding initial
angles in radians are:
$\phi_0\approx 2.20$
(black), $\phi_0\approx 2.44$ (grey), and $\phi_0\approx 2.68$
(light grey). Notice that in the black curve, e. g., we have
normal and reverse instantaneous motions depending on the instant
we record the velocity $\dot{\phi}$. The results presented in Fig.
\ref{fig2} refer to the global behavior of $\phi$.

\begin{figure}
\includegraphics[width=4.5cm,angle=-90]{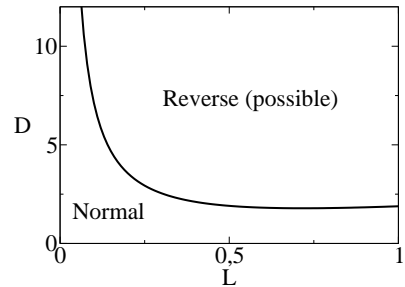}
\caption{Configurations located above the curve may develop
reverse rotations.
Below the curve only normal rotation is possible.}
\label{fig4}
\end{figure}

It is possible
to establish in a more precise way which initial configurations
lead to reverse rotations. First we note that the range of initial
angles that lead to reverse behavior is bounded by the
intersections of the synchronization line (\ref{rr}) and the
initial condition curve (\ref{cc}). These boundary angles are
given by
\begin{equation}
\cos^2 \phi_0^{(b)}= \frac{I_P^2}{(mdl\omega)^2}
\left(\frac{2\tilde{\cal E}}{I_P}-\frac{2mdl\omega^2}{I_P}-\omega^2 \right)\;.
\label{rr2}
\end{equation}
Since $0 \le \cos^2 \phi_0^{(b)} \le1$, in order to get reverse
rotations we must have
\begin{equation}
2mdl+I_P< \frac{2 \tilde{\cal E}}{\omega^2} < \frac{1}{I_P}(I_P+mdl)^2\;,
\label{rr3}
\end{equation}
where we have a condition imposed on $\tilde{\cal E}=\tilde{\cal
E}(m,d,l,I_P,\omega)$. Let us analyze the limiting cases: $\frac{2
\tilde{\cal E}}{\omega^2}=2mdl+I_P$ and $\frac{2 \tilde{\cal
E}}{\omega^2}=\frac{1}{I_P}(I_P+mdl)^2$. Substituting these
expressions in the transcendental equation (\ref{trans}) we obtain
relations involving a {\it single} parameter, $K\left(\sqrt{\frac{2
\sigma}{\sigma+1}} \right)=\frac{\pi}{2}\sqrt{\sigma+1}$ and
$K\left(\frac{2\sqrt{2 \sigma}}{\sigma+2}
\right)=\frac{\pi}{4}(\sigma+2)$, respectively, with
$\sigma=2mdl/I_P$. The first equation has only the solution
$\sigma=0$, which is physically trivial, implying that the left-hand side
of the inequality (\ref{rr3}) is always fulfilled. The second
equation, besides $\sigma=0$, presents the solution
$\tilde{\sigma} \approx 2.523$. Thus initial configurations
obeying $2mdl/I_P=\tilde{\sigma}$, separate regions where reverse
rotation is possible from regions where only normal rotations can
occur. This condition involves only the relative scales $D=d/R \in
(0,\infty)$ and $L=l/R \in (0,1)$ and, in the case of a disk, reads
\begin{equation}
D=0.631\,L^{-1}+1.261 \, L\;.
\label{universal}
\end{equation}
The above result is universal, in the sense that it is valid for
any $m$ and $\omega$, and is independent from the absolute scales
$R$, $l$, and $d$. As indicated in Fig. \ref{fig4}, initial
geometrical configurations located below the curve
(\ref{universal}) always lead to normal behavior, while
configurations above it may enable reverse rotations, depending on
the initial angle $\phi_0$ (the precise values of $\phi_0$ are
given by Eq. (\ref{rr2})). We also note that there is a value of
$d$ below which no reverse rotation occur. It is given by
$D_{min}=\tilde{\sigma}/\sqrt{2}$, that is, $d_{min}=1.784R$. For
$L>0.5$, variations in this parameter produce virtually no change in $D$,
which becomes the only relevant parameter to define the
possible regimes of the system (see the ``plateau'' in Fig.
\ref{fig4}).

We stress that the obtained results are very general since we used the
particular form of $I_P$ only to obtain Eq. (\ref{universal}).
For an arbitrary RB we have
$I_P=ml^2+\gamma m {\cal R}^2$,
where $\gamma$ is a number and ${\cal R}$ is a characteristic scale.
For the disk we have $\gamma=1/2$ and ${\cal R}=R$, while for a rectangular
plate
of sides $a$ and $b$ we have $\gamma=1/12$ and ${\cal R}=\sqrt{a^2+b^2}$.
The general form of Eq. (\ref{universal}) is
\begin{equation}
D=\frac{ \tilde{\sigma}}{2}(\gamma\, L^{-1}+ L)\;,
\label{universal2}
\end{equation}
with $D=d/{\cal R}$ and $L=l/{\cal R}$, for ${\cal R} \ne 0$.
We note that in the ``degenerate'' case of a
point mass connected by a massless rod to the pivotal point, obtained
from the disk by
taking $R=0$, we get $\sigma=2d/l$ and the reverse condition becomes
a one-parameter relation $d/l>1.261$.

We investigated reverse rotations in the circularly-driven motion of a RB
whose intrinsic angular degree of
freedom was shown to evolve according to a combination of pendular and uniform
motions. This enabled the complete determination of the initial configurations
that lead to reverse behavior (Eq. (\ref{rr3})). In addition a scale free,
purely geometrical relation, defining the regions were reverse
rotation is possible was obtained (Eq. (\ref{universal2})). The
effects of friction, assumed to be negligible in this work,
may play an interesting role. The friction generated in the small
contact area connecting the thin rod and the RB gives an extra
torque (but no net force). This may be the only friction in the
problem if we assume that the RB is kept in the horizontal
position solely by the rod. It may also be of interest to consider
the system immersed in a viscous fluid. This would make our
apparatus very similar to a bioreactor for tissue growth
\cite{bio}. It seems that the influence of the regime of intrinsic rotation of the
tissue construct (our disk) is yet to be analyzed. These pending
issues are presently under investigation.


\acknowledgments The author thanks G. L. Vasconcelos, M. A. F. Gomes,
and P. H. Figueiredo for helpful
discussions. J. A. Miranda is acknowledged for a
critical reading of the manuscript. This work was partially
supported by the Brazilian agencies CNPq and FACEPE (DCR
0029-1.05/06 and APQ 0800-1.05/06).

\end{document}